\documentclass[copyright]{eptcs}
\providecommand{\event}{WWV 2015} 
\usepackage{breakurl}             
\usepackage{graphicx}
\usepackage{comment}
\usepackage{color}
\usepackage{soul} 
\usepackage{color} 
\title{Domain-specific queries and Web search personalization: some investigations\thanks{This research has been partially funded by the Registro.it project
\textit{MIB}, My Information Bubble.}}
\author{Van~Tien~Hoang
\institute{IMT Lucca, Italy}
\email{vantien.hoang@imtlucca.it}
\and
Angelo~Spognardi
\institute{IIT-CNR, Pisa, Italy}
\email{a.spognardi@iit.cnr.it}
\and
Francesco~Tiezzi
\institute{University of Camerino, Italy}
\email{francesco.tiezzi@unicam.it}
\and
Marinella~Petrocchi
\institute{IIT-CNR, Pisa, Italy}
\email{m.petrocchi@iit.cnr.it}
\and
Rocco De Nicola
\institute{IMT Lucca, Italy}
\email{rocco.denicola@imtlucca.it}
}

\def\titlerunning{Domain-specific queries and Web search personalization}
\def\authorrunning{V. T. Hoang et al.}
\begin{document}
\maketitle


\begin{abstract}

Major search engines deploy personalized Web results to enhance users' experience, by showing them data supposed to be relevant to their interests. Even if this process may bring benefits to users while browsing, it also raises concerns on the selection of the search results. In particular, users may be unknowingly trapped by search engines in protective information bubbles, called ``filter bubbles'', which can have the undesired effect of separating users from information that does not fit their preferences. This paper moves from early results on quantification of personalization over Google search query results. Inspired by previous works, we have carried out some experiments consisting of search queries performed by a battery of Google accounts with differently prepared profiles. Matching query results, we quantify the level of personalization, according to topics of the queries and the profile of the accounts. This work reports  initial results and it is a first step a for more extensive investigation to measure Web search personalization.
\end{abstract}

\section{Introduction}
Traditional Web search services use nature of requests aside from user personal preferences and search intents.  They combine many elements to guess users' needs in order to provide better fitting data and improve their Web experience. The results provided for a query may be influenced  by individual factors and personal contexts, like long-term search history~\cite{Matthijs2011}, click-through entropy on search result links~\cite{Dou2007}, search sessions~\cite{Ustinovskiy2013} and users' bookmarks~\cite{Hu2008}.

Search providers want to supply users with more relevant data by personalizing the query results. This is good for users at first, but may also have undesired effects. Indeed, users retrieve richer data about specific domains that search providers think they are looking for. However, this creates a trapping effect,  so-called ``filter bubbles''~\cite{filter2011} effect, in which users can only reach information that search engines tailor for. Users may not be aware of these filters and, if they want to explicitly change how search engines categorize their interests, they still may not be able do that. 
%

To understand how filter bubbles take shape for a given Web user, first it is necessary to assess the level of personalization of results provided to that user by search services.
Previous work has evaluated personalization of Web results, exploiting user's features and history information (see, e.g., ~\cite{Matthijs2011,Nanda2014,Makvana2014}). In this paper, we aim at understanding the level of personalization proposed by a commercial search engine ---Google--- to its users, in comparison with non-Google users. A Google user is one that has a Google account and logs into her Google account before performing any search activity. A user who accesses Google search service without logging in is called non-Google (or \textit{vanilla}) user.

We check the effects that logging into Google accounts have on search results by building Google user profiles based on search histories and queries' topics. After that, we performed experiments with different Google user accounts settings on local network and then evaluated the effects of Web search personalization, comparing to non-Google users.

The paper is organized as follows. In Section~\ref{sec:backgr-motiv-relat}, we present previous related work and we address motivations for  and contributions of the current paper. Section~\ref{sec:meth-exper-sett} shows the methodology and the experiments. Section~\ref{sec:results} reports details and comments on experiment results. Finally, we conclude in Section~\ref{sec:conclusion}.

\section{Background, Motivation, and Related Work}\label{sec:backgr-motiv-relat}
We briefly introduce background notions used in the rest of the paper. Then, we discuss related work in the area and we give motivations for the current work. 

A (Web) search personalization consists of a set of modifications performed over the set and order of query search results, when considering a search query performed by a specific user. 

We use non-Google user or vanilla user interchangeably to refer to a user that searches on the Google website without logging in; otherwise, she is a Google user. 
A query is made of keywords that the user inputs into the search engine. Each query may be referred to a certain category, which, in its turn, is linked to users' interests. A Google user has a profile when her interests are in specific categories.  



Finally, to measure the level of personalization in search results, users search for queries taken from  a list of keywords, called test keywords list. This list contains terms that fit into different interests.\\

Web search personalization has attracted many works in recent years. We can define two main approaches for the personalization, considering the type of features used to realize it: that based on user-centric features and that based on history-centric features. The former exploits all user relevant data such as gender, location and click history. The user-centric approach requires to store a large amount of data and can consist of information not available to everyone, i.e., server logs, personal user data and so on). The latter approach utilizes data related to the behavior of the user over time, in both short-term and long-term period (i.e., Web browsing and searching history), as well as in very short-term (browsing session)~\cite{Yue2014}.

Nanda \textit{et al.}~\cite{Nanda2014} 
created an ontology-based profile for users by building a hierarchy of topic trees from Web (e.g., from Open directory project and Wikipedia), then they combined it with explicit user interests (users provided bookmark links and some keywords they associated with a topic). 
Profiles then evolve through collaborative filtering using the k-nearest neighbor-based algorithm by terms between similar users. 
Later, the authors proposed a technique to re-rank the results from search engines according to their relevance to a user, based on her learned profile.
Matthijs and Radlinski~\cite{Matthijs2011} also used long-term search history 
to develop models of user's interests and used those models to re-rank Web results. 

Yury \textit{et al.}~\cite{Ustinovskiy2013} considered short-term context by exploiting browsing history and first queries of search sessions. A search session is a series of intent-related users queries issued to a search engine. The authors 
predicted which Web result links are clicked by modeling features from search session context like queries, click-through and browsing. The links were then categorized by a hierarchical ontology structure, based on Open Directory Project (ODP)\footnote{http://www.dmoz.org}. Finally, they applied a re-ranking function according to previous prediction model to create a personalized Web results.
Ryen \textit{et al.}~\cite{white} also studied short-term context, current session and query. They proposed to combine and weight the context of each query to predict short-term interests of users.

Makvana \textit{et al.}~\cite{Makvana2014}, as opposed to client-side history, analyzed Web logs from servers. They identified related search terms for a particular user from previous searches history, and used these related terms to clarify her search intent for ambiguous queries. To do that, the queries were expanded by adding other related terms to them. For example, the ambiguous query ``apple" was transformed into either ``apple fruit" or ``apple ipod" depending on user's search history.  
Moreover, they processed user's search query results and used Vector Space Model (VSM) to generate user interest values on the links, and produced new ranks of links.

Hu and Chan~\cite{Hu2008} proposed a scoring function that uses term characteristics and image term characteristics to score a term that matches users profile, which is learned from users bookmarks. Web results are then customized according to the scoring function.
Authors in \cite{yu2010mining} also applied a user-centric approach, analyzed short-term query context, and user context like clicks and links to apply in personalized search of session. 
Other works combine both user-centric and history-centric approaches. As an example, Yue \textit{et al.}~\cite{Yue2014} used short-term and session-based history, collected in 3 months, to generate a probabilistic model and a click-through model to customize search results for users. 
Mikhail and Matthew~\cite{Mikhail2011} 
used a similar approach when applied probability to customize what should be shown to users.

The most inspiring study for our work is that of Hannak \textit{et al.}~\cite{Hannak}, in which the authors propose methods and ideas to measure personalization of Google search engine. They evaluated the effects on query results of many factors of a user profile (i.e., age, gender, income, location), system related data (i.e., browser, operating system) and server-side technology (i.e., cookies). They found out that Google has low level of personalization of results in general, and it is based mainly on geo-localization.   
Stimulated by the unexpected low level of personalization of their results, which contrasts the assumptions on personalization  
at the basis of the filter bubble thesis, we further investigate in our work on personalization of Google search query results, by performing other kind of experiments.

Our main objective is indeed to better understand the filter bubble created by a commercial search engine like Google. The first step is to acknowledge how those bubbles are formed, namely which features of users and environment the search engine relies on to personalize the results. Then, we try to identify which factors Google uses to fill the profiles and, finally, quantifying the level of personalization on search query results, in terms of set and order of results. Inspired by previous works, we expand existing investigations by trying to train Google profiles on particular topics and we compare the results of search queries by the accounts holding different trained profiles. 

%
%

We believe the study presented hereafter  is a starting step and provides additional data to further research.

\section{Methodology and Experiments Setting}\label{sec:meth-exper-sett}
\label{sec:exp}
We describe in this section the methodology followed for quantifying the current level of personalization of Google Web searches and we illustrate the settings of the conducted experiments.   

\subsection{Methodology}
In our study, we quantify personalization level by measuring differences in search results between users when they query for the same keywords. 
The differences can be in the relative ranking of search query results and in the results themselves. We also check if there are differences between a Google user - profiled by means of the queries - and a plain Google user, with a plain profile (or  a non-Google user). 


In order to compare two search query results, we only considered the normal query reults, ignoring the Image and News boxes. As evaluation metrics, we employed the Jaccard Index and the Edit Distance, also used by Hannak \textit{et al.}~\cite{Hannak}. Jaccard Index of two sets A and B is defined as ${|A\cap B|}\,/\,{|A \cup B|}$. It ranges from 0 to 1 and  measures the differences between the elements of two sets, without considering their order. A Jaccard Index of 0 means that two sets are disjoint, 1 means that two sets are identical. Edit Distance, instead, measures how many operations (insertion, deletion, substitution or swap) are needed to transform one list into another. For example, if user $u_i$ receives search result set [A, B, C, D] and user $u_j$ gets [D, C] for the same query, then the Edit Distance is 3 (two insertion and one swap).

We describe in the following our experiments. First, we created multiple Google accounts with different values in gender, age and location. The accounts were made by normally registering via Google website. Then, we logged in all of them and executed the same queries simultaneously, on a single IP address in different browser sessions. We used a single IP address to eliminate the potential noise due to a difference in geographic locations, which is one of the most significant elements affecting search results, see~\cite{Hannak}. At the same time, we used many other vanilla users to execute exactly the same queries. Once the experiments had ended, we compared result pages in terms of Jacquard Index and Edit Distance.

In a second phase, we have built Google accounts with narrower interests, labeling them as profiled Google accounts. These accounts were normally registered from Google and they only searched for words in lists of keywords  (the \textit{training keyword list}) about specific and narrow domains. 
In particular, we have considered the football domain and built a list of keywords composed by football players, coaches, staff people of football clubs (e.g., ``AC Milan coach Nereo Rocco", ``Fly Emirates" - a sponsor of AC Milan, ``football Inter Milan Rodrigo Palacio" - a player of Inter Milan). 
We also set up another keywords list (the \textit{test keyword list}) containing deliberately vague terms, again related to football teams, such as ``next match" and ``home stadium". Then, we used each user to execute search queries in the same domain, but with different training keyword lists at the same time. Later, we evaluate whether those differently built accounts received results influenced by their history of queries. In order to do that, we asked them performing queries on the test keywords list. Moreover, we also compared their results with those of users that did not search for the training keywords list.

Summarizing, we performed the following experiments:
\begin{itemize}
\item Experiment 1: 
We have tried to build profiled accounts, letting each of them search in different domains, in a training phase. Then, they searched on a test keyword list, in a domain different from the training ones.  By comparing web results on the test queries, it is possible to measure the personalization effect of the training queries.
\item Experiment 2: 
We have tried to build profiled accounts, letting them search for keywords in a narrow domain. Then, they searched on the test keyword list (a less narrow domain) to evaluate if some relevant differences emerge. As before, by comparing web results on the test queries, it is possible to measure the personalization effect of the training queries.
\item Experiment 3: 
We repeated experiment 2 with different query composition, in order to evaluate if and how this affects the user profile and its Google exploitation.
\item Experiment 4: We have tested how the test keyword results change when a Google user perform massive queries on narrow topics for a long time.
\end{itemize}

\subsection{Experiment settings}

Our experiment settings and tools are strongly inspired by and adapted from original work of Hannak \textit{et al.}~\cite{Hannak}. All experiments run on 
PhantomJS\footnote{http://www.phantomjs.org}, a headless web-browser based on Webkit. Basically, it is a full browser with a Java-script engine as well as a modern web navigator software, but without a user interface. We acknowledge that Google uses many servers with several IP addresses. Since each server could have different index databases according to its location, to eliminate the risk of noise, we fixed one IP address for the Google server in the \textit{host} file of the operating system. In this way, we expected that all search requests are  routed to a single server address of Google. 
Using PhantomJS, each account is logged in right after its creation, to mimic the behavior of a real user. When performing a search, all accounts operate at the same time, with an interval of 11 minutes between two searches, to avoid carry-over effect~\cite{Hannak}. The carry-over effect is a phenomenon that happens when users perform two sequential queries A and B: the results of query B are influenced by previous search for A. For example, if one - not necessarily profiled -  searches for ``python" and after searches for ``programming language", it is likely to see results relevant to the Python language.
Details on the experiments settings and results is available online at \url{https://sites.google.com/a/imtlucca.it/wwv2015/}.  

\section{Results}\label{sec:results}

In this section, we comment on the results of the experiments described in Section~\ref{sec:exp}.
 
\subsection{Experiment 1}
In this experiment, we check how the details in users' profiles influence their search results. In particular, we have created three Google accounts and we have let them search keywords in the following categories: 
    \begin{itemize}
        \item Account 1, with training keywords from shoes and baseball categories;
        \item Account 2, with training keywords from drinks, foods, and retail brands;
        \item Account 3, with training keywords from politics, fashion, and shopping.
    \end{itemize}
All Google accounts in our experiments have been manually enrolled from Gmail registration page.
The keywords are  what the accounts are querying for and the queries are grouped by their semantic meanings. In particular, we have used keywords from Google Trends (August 2014, US), which has categorized popular search queries in corresponding categories. We let all accounts search for those keywords. After this training phase, they search on test keywords list.
    \begin{table}[ht]
         \begin{center}
         \scalebox{0.8}{
                \begin{tabular}{ll}
                \hline
                Number of search terms & Number of test terms                               \\ \hline
                \multicolumn{1}{c}{140} & \multicolumn{1}{c}{19, running time \textless 24h} \\ \hline
                \end{tabular}
            }
            \end{center}
        \caption{Setting for Experiment 1}
        \label{ex1}
    \end{table}
    \begin{figure}[h]
        \centering
        \includegraphics[width=0.9\textwidth]{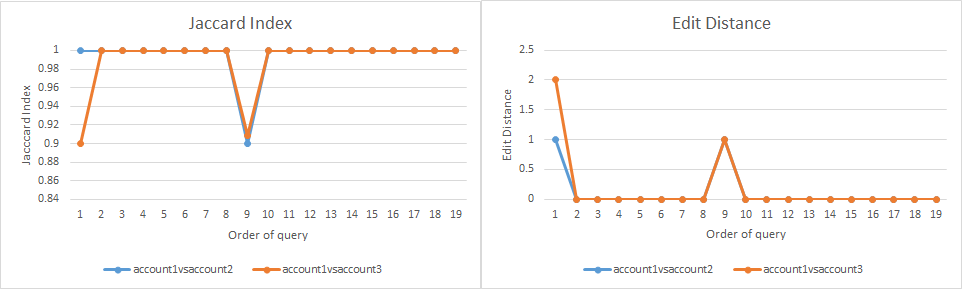}
        \label{ex11}
        
        \caption{Jaccard Index (Left) and Edit Distance (Right) for Experiment 1.}
        \vspace*{-.3cm}
    \end{figure}
    

In the experiment, 4 out of 38 test queries produced different results (account vs account). 
When we check the natural language meaning of the different ones, we were unable to find a clear connection between them. For example, with test keyword ``Plato" searched by Google accounts 1 and 2, Jaccard Index is 0.9, meaning 90\% of the results are the same (we checked the first 10 results shown by Google). The only difference between them is 1 web link about ``PLATO 2.0 - A space agency" which has no correlated meaning to the previous training searches of both the users (Fig.1). 
Quite obviously, checking only the first ten results provided by Google do not give us the capability to conclude with a final claim on the fact that  topics in previous queries interfere with results of the test queries. What we can assert is that, for previous queries belonging to the categories in the list shown above, the three accounts under investigation obtain no significant differences in the first ten results on the test queries. 

\subsection{Experiment 2}
In this experiment, we have tried to train Google accounts to emulate the fact that they have a strong interest in football. We chose two football teams (the Italian AC Milan and Inter Milan) and we collected keywords from their official websites. Those keywords are directly related to the corresponding football clubs, such as history of achievements, captains, coaches, leader boards, management staff and players. All the search query terms  exist in the football club official websites.
    \begin{table}[h]
             \begin{center}
             \scalebox{0.8}{
                    \begin{tabular}{ll}
                    \hline
                    Number of search terms & Number of test terms                               \\ \hline
                    \multicolumn{1}{c}{134} & \multicolumn{1}{c}{26} \\ \hline
                    \end{tabular}
                }
                \end{center}
        \vspace*{-.5cm}                
            \caption{Setting for Experiment 2.}
            \label{ex2}
        \end{table}

After calculating Jaccard Index and Edit Distance, we found that there were 6 out of 25  test queries yielding different results (see Fig.2). 
However, those different results were not connected to the two football clubs under investigation (i.e., when searching for ``next match", the different results were not about next match of AC Milan or Inter Milan teams). Again, such an outcome let us to assert that, even if one account has searched for very narrow domains in its past activity over Google, apparently that activity does not influence new results when searching for vague, still related, domains. However, further investigation is needed for more rigorous claims on the matter. 

\subsection{Experiment 3}
Experiment 3 was a revised version of Experiment 2. In Experiment 3, the search query had the  form ``football" + [club's name] + keyword. 
    \begin{figure}[h]
      \centering
       \includegraphics[width=1\textwidth]{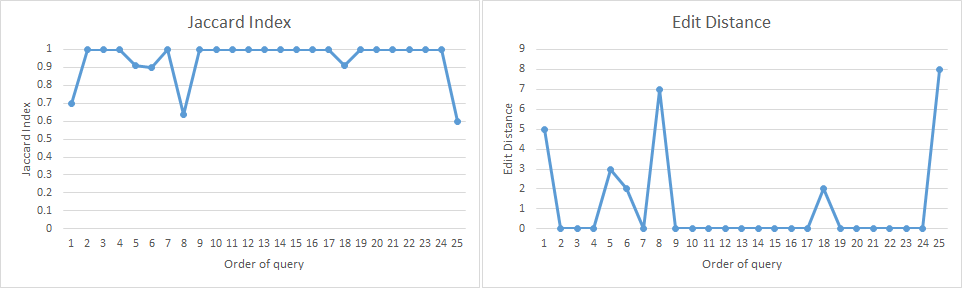}
       \label{ex21}
       \vspace*{-.6cm}
       \caption{Jaccard Index (Left) and Edit Distance (Right) for Experiment 2.}
       
    \end{figure}
    \begin{figure}[h]
      \centering
       \includegraphics[width=1\textwidth]{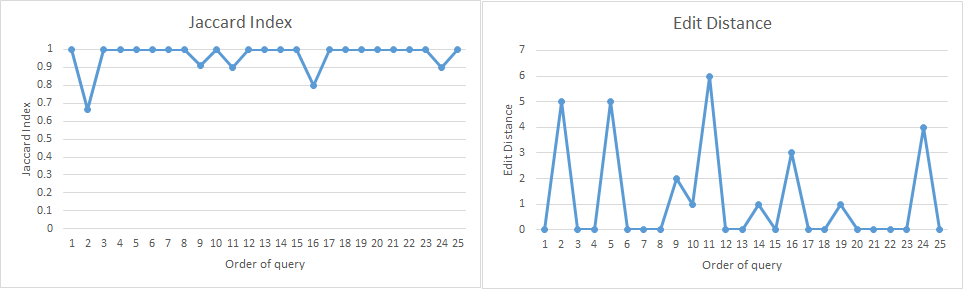}
       \label{fig:ex22}
       \caption{Jaccard Index (Left) and Edit Distance (Right) for Experiment 3.}
    \end{figure}
As shown in Fig.~3, the order of search results among the accounts under investigation shows more differences  than the previous experiment. This paves the way for further investigation. 
\subsection{Experiment 4}
\begin{table}[!h]
    \begin{center}
    \scalebox{0.8}{
        \begin{tabular}{cc}
        \hline
        Search topics                                                                                                                                                                                                                                                            & Notes                       \\ \hline
        \begin{tabular}[c]{@{}c@{}}layers, captains, technical staffs, \\ coaching staffs, management history,\\  administrative staffs, \\ name of championship (i.e. won cups),\\ list of sponsors (technical and/or official),\\ list of captains in history\end{tabular} & There are repeated keywords \\ \hline
        \end{tabular}
    }
    \end{center}
\vspace*{-.5cm}    
     \caption{Keyword setting for Experiment 4.}
    \label{my-label}
\vspace*{-.3cm}      
\end{table}
We have selected 400 keywords (topics in Table 3) about one football team and we have created 2 new Google users to continuously perform searches  for 72 hours. After a massive number of search queries in a narrow topic, we expected the results from test keywords to be significantly different. Instead, both Jaccard Index and Edit Distance showed a limited variation.


\section{Conclusions}\label{sec:conclusion}

Web search personalization services are available in many web search engines,  like Google and Yahoo (consider, for example, the ``intelligent personal assistant"  Google Now, available for Google search on Android and Chrome). In this paper, we started from previous work in the area and continued to investigate the level of Web search personalization on Google. We proposed a series of experiments settings that may be useful and inspiring for understanding how characteristics of a set of accounts (mainly, previous search activities of those accounts) may influence further searches. The obtained results do not show a marked level of Google personalization, at least regarding the kind of queries we have chosen to perform (both training and test queries)
 and the number of pages results we have analyzed. 
 However, the methodology and these initial results constitute an initial step. As an example, in this paper we have simulated many computers within a single local network. The results could be biased because of the single IP address. We are carrying out the same experiments on the cloud (Amazon Cloud) to see how geographic location could affect search results. We also plan to study different techniques and use other initial queries and test queries.

\nocite{*}
\subsection*{Acknowledgements}
We thank the anonymous reviewers for their support to check over our
experiment results. We also thank the WWV workshop attendants for the
fruitful discussion that followed the presentation of our work.
\bibliographystyle{eptcs} \bibliography{generic-reconstr}
\end{document}